\documentstyle[amssymb,aps]{revtex}

\newtheorem{theorem}{Theorem}
\newtheorem{acknowledgement}[theorem]{Acknowledgement}

\begin{document}
\title{Alternative to $R_{\lambda }$-scaling of Small-Scale Turbulence Statistics}
\author{Reginald J. Hill$^{a}$}
\address{National Oceanic and Atmospheric Administration, Environmental Technology\\
Laboratory, 325 Broadway, Boulder CO 80305-3328\\
(PACS 47.27.Gs, 47.27.Jv)\\
$^{a}$tel:3034976565, fax:3034976181, Reginald.J.Hill@noaa.gov}
\date{\today }
\maketitle

\begin{abstract}
Traditionally, trends of universal turbulence statistics are presented
versus $R_{\lambda }$, which is the Reynolds number based on Taylor's scale $%
\lambda $ and the root-mean-squared (rms) velocity $u_{rms}$. \ $\lambda $
and $u_{rms}$, and hence $R_{\lambda }$,\ do not have the attribute of
universality. \ The ratio of rms fluid-particle acceleration to rms viscous
acceleration, $R_{{\bf a}}$, is an alternative to $R_{\lambda }$ that has
the advantage of being determined by the small scales of turbulence. \ This
ratio has the following attributes: $R_{{\bf a}}$ is a Reynolds number, it
is composed of statistics of the small scales of turbulence, can be
evaluated with single-wire hot-wire anemometry, and like $R_{\lambda }$, can
be partially evaluated by means of flow similarity. \ For isotropic
turbulence the relationship between $R_{{\bf a}}$ and $R_{\lambda }$ is
given. \ Graphs of velocity derivative flatness measured in anisotropic
turbulence driven by counter rotating blades have different appearances
depending on whether $R_{{\bf a}}$ or $R_{\lambda }$\ is the abscissa.
\end{abstract}

\section{INTRODUCTION}

\qquad Reynolds\cite{Reynolds} sought, from the Navier-Stokes equation,
``the dependence of the character of motion on a relation between the
dimensional properties and the external circumstances of motion.'' \
Assuming that the motion depends on a single velocity scale $U$\ and length
scale $c$, Reynolds found that the accelerations are of two distinct types
and thereby deduced that the relevant solution of the Navier-Stokes equation
``would show the birth of eddies to depend on some definite value of $c\rho
U/\mu $,''\cite{Reynolds} where $\rho $ is the mass density of the fluid and 
$\mu $\ is the coefficient of viscosity. \ Reynolds performed exhaustive
experiments that demonstrated his deduction, as well as experiments on the
stabilization of fluctuating flow.\cite{Reynolds} \ He discovered the sudden
onset of flow instability\cite{Reynolds}. \ The Navier-Stokes equation is $%
{\bf a}=\partial {\bf u}/\partial t+{\bf u}\cdot {\bf \nabla u}=-{\bf \nabla 
}p+\nu \nabla ^{2}{\bf u}$, where $p$ is pressure divided by $\rho $, $\nu
=\mu /\rho $ is kinematic viscosity, ${\bf u}$ is the velocity vector, and $%
{\bf a}$ is the acceleration. \ Batchelor \cite{Batch70} discussed (in his
Sec. 4.7) the interpretation of the Reynolds number as a measure of ${\bf a}$%
\ relative to the viscous term $\nu \nabla ^{2}{\bf u}$. \ He noted that the
balance of the Navier-Stokes equation can also be parameterized in terms of
the relative magnitudes of ${\bf \nabla }p$\ and $\nu \nabla ^{2}{\bf u}$. \
The latter parameterization does not technically lead to a Reynolds number,
but it will be shown that the two parameterizations become equivalent at
large Reynolds numbers. \ We therefore introduce the two ratios: \ 
\begin{equation}
R_{{\bf a}}\equiv \left\langle {\bf a}\cdot {\bf a}\right\rangle
^{1/2}/\left\langle \nu ^{2}\left( \nabla ^{2}{\bf u}\right) {\bf \cdot }%
\left( \nabla ^{2}{\bf u}\right) \right\rangle ^{1/2}\text{ \ and \ }R_{{\bf %
\nabla }p}\equiv \left\langle {\bf \nabla }p\cdot {\bf \nabla }%
p\right\rangle ^{1/2}/\left\langle \nu ^{2}\left( \nabla ^{2}{\bf u}\right) 
{\bf \cdot }\left( \nabla ^{2}{\bf u}\right) \right\rangle ^{1/2},
\label{definitions}
\end{equation}
where $\left\langle \cdot \right\rangle $\ denotes an average. $\ R_{{\bf a}%
} $ is a Reynolds number in Batchelor's concept, and $R_{{\bf \nabla }p}$\
is a parameterization of the relative magnitudes of ${\bf \nabla }p$\ and $%
\nu \nabla ^{2}{\bf u}$.

\qquad To paraphrase Nelkin's\cite{Nelkin94} description of Reynolds number
scaling: \ if two turbulent flows have the same geometry and the same
Reynolds number, then their statistics, when appropriately scaled, should be
equal. \ A statistic of the small scales of turbulence is an average of
quantities that contain only products of differences, such as two-point
velocity difference or derivatives of velocity. \ Universality of the small
scales of turbulence is the hypothesis that statistics of the small scales,
when appropriately scaled, should become equal as Reynolds number increases 
\cite{Nelkin94}\cite{SreenAnt}; that is, the flow geometry becomes
negligible in the limit that the Reynolds number is infinite. \ Discovering
the appropriate scaling that results in universality is the topic of a vast
amount of research\cite{Nelkin94}\cite{SreenAnt} and will not be pursued
here. \ The relevance of universality to real turbulent flows is discussed
by Nelkin\cite{Nelkin94} and Sreenivasan and Antonia\cite{SreenAnt}.

\qquad The Reynolds number based on\ the root-mean-square (rms) of the
longitudinal-velocity component $u_{rms}\equiv \left\langle
u_{1}^{2}\right\rangle ^{1/2}$ and Taylor's length scale $\lambda $ is $%
R_{\lambda }\equiv u_{rms}\lambda /\nu $, where $\nu $ is kinematic
viscosity, and $\lambda \equiv u_{rms}/\left\langle \left( \partial
u_{1}/\partial x_{1}\right) ^{2}\right\rangle ^{1/2}$. \ Here, $u_{1}$ and $%
x_{1}$ are the components of velocity and spatial coordinate in the
direction of the $1$-axis. \ For decades,$\ R_{\lambda }$ has been used as
the abscissa for presenting statistics that are believed to be universal
aspects of small-scale turbulence (such as velocity derivative statistics
normalized by powers of $\left\langle \left( \partial u_{1}/\partial
x_{1}\right) ^{2}\right\rangle $). \ The observed trends as $R_{\lambda }$\
increases are an often-sought quantification of scaling universality. \ $%
R_{\lambda }$ has the advantage of being easily measured because it requires
only measurement of $u_{1}$ (which yields $\partial u_{1}/\partial x_{1}$\
by means of Taylor's hypothesis); that measurement can be obtained with a
single hot-wire anemometer. \ Alternatively, flow similarity can be used to
estimate the energy dissipation rate $\varepsilon $, and by substituting the
local-isotropy relationship that $\varepsilon =15\nu \left\langle \left(
\partial u_{1}/\partial x_{1}\right) ^{2}\right\rangle $, $R_{\lambda }$ can
be obtained from $R_{\lambda }=u_{rms}^{2}/\left( \varepsilon \nu /15\right)
^{1/2}$. \ Because $R_{\lambda }$\ depends on $u_{rms}$, it depends on
large-scale geometry of the flow. \ Nelkin\cite{Nelkin94} discussed the
nonuniversal attributes of $R_{\lambda }$. \ As a result of the
nonuniversality of $R_{\lambda }$, statistics of the small scales, e.g.,
normalized derivative moments, when graphed with $R_{\lambda }$\ on the
abscissa, can have different curves corresponding to dissimilar flows.

\section{Alternative}

\qquad In addition to graphing such statistics with $R_{\lambda }$ on the
abscissa, it would seem advantageous to use a quantity on the abscissa that
is solely a property of the small scales of turbulence. \ That advantage has
long been recognized.\cite{Wyntenn}\cite{VanAttaAnton}\cite{AntChamSaty} \
Here, we consider the alternative $R_{{\bf a}}$ defined in (\ref{definitions}%
), and determine how it can be measured with an instrument no more complex
than a single-wire hot-wire anemometer. \ Because the intended application
is to statistical characteristics of the small scales, it is appropriate to
simplify $R_{{\bf a}}$ and $R_{\lambda }$ on the basis of local isotropy. \
Indeed, local isotropy is a precondition for universality.\cite{Nelkin94} 
\cite{SreenAnt} \ On this basis, $\left\langle {\bf a}\cdot {\bf a}%
\right\rangle =\left\langle {\bf \nabla }p\cdot {\bf \nabla }p\right\rangle
+\left\langle \nu ^{2}\left( \nabla ^{2}{\bf u}\right) {\bf \cdot }\left(
\nabla ^{2}{\bf u}\right) \right\rangle $\cite{ObukYaglom}\cite
{HillThoroddsen}; in which case 
\begin{equation}
R_{{\bf a}}=\sqrt{1+R_{{\bf \nabla }p}^{2}}\text{ and }R_{{\bf \nabla }p}=%
\sqrt{R_{{\bf a}}^{2}-1}.  \label{alternative}
\end{equation}
Because $\left\langle {\bf \nabla }p\cdot {\bf \nabla }p\right\rangle \gg
\left\langle \nu ^{2}\left( \nabla ^{2}{\bf u}\right) {\bf \cdot }\left(
\nabla ^{2}{\bf u}\right) \right\rangle $ at high Reynolds numbers\cite
{HillThoroddsen}, (\ref{alternative}) becomes $R_{{\bf a}}\simeq R_{{\bf %
\nabla }p}$.

\subsection{Menu for the variance of viscous acceleration}

\qquad The correlation of viscous acceleration is\cite{ObukYaglom} 
\begin{equation}
V_{ij}\left( {\bf r}\right) \equiv \left\langle \nu ^{2}\nabla _{{\bf x}%
}^{2}u_{i}\nabla _{{\bf x}^{\prime }}^{2}u_{i}^{\prime }\right\rangle =-%
\frac{\nu ^{2}}{2}\nabla _{{\bf r}}^{2}\nabla _{{\bf r}}^{2}D_{ij}\left( 
{\bf r}\right) ,  \label{Vii}
\end{equation}
where prime denotes evaluation at a point ${\bf x}^{\prime }$; ${\bf x}$\
and ${\bf x}^{\prime }$\ are independent variables; ${\bf r\equiv x-x}%
^{\prime }$; $\nabla _{{\bf r}}^{2}$ is the Laplacian operator in ${\bf r}$%
-space; the right-most expression in (\ref{Vii}) is obtained on the basis of
local homogeneity. \ Let $\varepsilon $\ denote the energy dissipation rate,
and $D_{ij}\left( {\bf r}\right) $ and $D_{ijk}\left( {\bf r}\right) $
denote the second- and third-order velocity structure functions. \ The
Navier-Stokes equation and local isotropy give\cite{Hill97} $\partial
_{t}D_{ij}\left( {\bf r}\right) +\partial _{r_{k}}D_{ijk}\left( {\bf r}%
\right) +\left( 4/3\right) \varepsilon =2\nu \nabla _{{\bf r}%
}^{2}D_{ij}\left( {\bf r}\right) $; applying the operator $-\left( \nu
/4\right) \nabla _{{\bf r}}^{2}$\ to that equation and comparison with (\ref
{Vii}) gives, 
\begin{equation}
V_{ij}\left( {\bf r}\right) =-\left( \nu /4\right) \nabla _{{\bf r}}^{2}%
\left[ \partial _{t}D_{ij}\left( {\bf r}\right) +\partial
_{r_{k}}D_{ijk}\left( {\bf r}\right) \right] .  \label{Moninseq}
\end{equation}
Derivative operators are abbreviated; e.g., $\partial _{t}=\partial
/\partial t$, $\partial _{r}^{2}=\partial ^{2}/\partial r^{2}$, etc. \
Summation is implied by repeated indexes. \ Performing the contraction of (%
\ref{Vii}) and (\ref{Moninseq}) such that the terms become functions of $r$
and $\nabla _{{\bf r}}^{2}\rightarrow \partial _{r}^{2}+\left( 2/r\right)
\partial _{r}$, we have, on the basis of local isotropy, \ 
\begin{eqnarray}
V_{ii}\left( r\right) &=&-\nu ^{2}\left( \partial _{r}^{4}+\frac{5}{r}%
\partial _{r}^{3}\right) D_{\beta \beta }\left( r\right) =-\frac{\nu ^{2}}{2}%
\left( r\partial _{r}^{5}+11\partial _{r}^{4}+\frac{24}{r}\partial
_{r}^{3}\right) D_{11}\left( r\right)  \nonumber \\
&=&-\partial _{t}\frac{\nu }{4}\left[ \partial _{r}^{2}+\frac{2}{r}\partial
_{r}\right] D_{ii}\left( r\right) +\frac{\nu }{2}\left[ -\frac{1}{r^{3}}%
D_{111}\left( r\right) +\left( -\partial _{r}^{3}-\frac{7}{r}\partial
_{r}^{2}-\frac{3}{r^{2}}\partial _{r}+\frac{6}{r^{3}}\right) D_{1\beta \beta
}\left( r\right) \right] .  \label{ViiD111}
\end{eqnarray}
Power-series expansion of the structure functions followed by
differentiation and taking the limit $r\rightarrow 0$ gives various formulas
for the viscous acceleration variance: 
\begin{eqnarray}
V_{ii}\left( 0\right) &=&\left\langle \nu ^{2}\left( \nabla ^{2}{\bf u}%
\right) {\bf \cdot }\left( \nabla ^{2}{\bf u}\right) \right\rangle =12\nu
^{2}\left\langle \left( \partial _{x_{1}}^{2}u_{\beta }\right)
^{2}\right\rangle =35\nu ^{2}\left\langle \left( \partial
_{x_{1}}^{2}u_{1}\right) ^{2}\right\rangle  \nonumber \\
&=&-\frac{1}{2}\partial _{t}\varepsilon -\frac{105}{4}\nu \left\langle
\left( \partial _{x_{1}}u_{\beta }\right) ^{2}\partial
_{x_{1}}u_{1}\right\rangle =-\frac{1}{2}\partial _{t}\varepsilon -\frac{35}{2%
}\nu \left\langle \left( \partial _{x_{1}}u_{1}\right) ^{3}\right\rangle ,
\label{Vii0}
\end{eqnarray}
where $\partial _{x_{1}}^{2}\equiv \partial ^{2}/\partial x_{1}^{2}$. \ The
term $-\left( 1/2\right) \partial _{t}\varepsilon $, which vanishes for
local stationarity, is included above, but was neglected by Ref.\cite
{HillThoroddsen}. \ The enstrophy generation rate can be written as:\cite
{TsinoKitDrac} \ $\left( -35/2\right) \left\langle \left( \partial
_{x_{1}}u_{1}\right) ^{3}\right\rangle =\left\langle \omega _{i}\omega
_{j}s_{ij}\right\rangle =\left( -4/3\right) \left\langle
s_{ij}s_{jk}s_{ki}\right\rangle $; comparing this with the right-most
expression in (\ref{Vii0}) gives expressions that can be evaluated by DNS or
multi-wire anemometers\cite{TsinoKitDrac}\cite{ZhouAnt}: 
\begin{equation}
V_{ii}\left( 0\right) =\left\langle \nu ^{2}\left( \nabla ^{2}{\bf u}\right) 
{\bf \cdot }\left( \nabla ^{2}{\bf u}\right) \right\rangle =-\frac{1}{2}%
\partial _{t}\varepsilon +\nu \left\langle \omega _{i}\omega
_{j}s_{ij}\right\rangle =-\frac{1}{2}\partial _{t}\varepsilon -\frac{4}{3}%
\nu \left\langle s_{ij}s_{jk}s_{ki}\right\rangle ,
\end{equation}
where $\omega _{i}$\ is vorticity vector and $s_{ij}$\ is rate-of-strain
tensor. \ The velocity-derivative skewness is 
\begin{equation}
S\equiv \left\langle \left( \partial _{x_{1}}u_{1}\right) ^{3}\right\rangle
/\left\langle \left( \partial _{x_{1}}u_{1}\right) ^{2}\right\rangle
^{3/2}=\left\langle \left( \partial _{x_{1}}u_{1}\right) ^{3}\right\rangle
/\left\langle \varepsilon /15\nu \right\rangle ^{3/2},  \label{skewness}
\end{equation}
such that (\ref{Vii0}) gives 
\begin{equation}
\left\langle \nu ^{2}\left( \nabla ^{2}{\bf u}\right) {\bf \cdot }\left(
\nabla ^{2}{\bf u}\right) \right\rangle =-\frac{1}{2}\partial
_{t}\varepsilon +0.3\varepsilon ^{3/2}\nu ^{-1/2}\left| S\right| .
\label{similarity}
\end{equation}
The variation of $\left| S\right| $\ with $R_{\lambda }$\ is known. \
Herring and Kerr\cite{HerrKerr}\cite{Kerr85} show $\left| S\right| $\
increasing from $0.074$ at\ $R_{\lambda }=0.46$ to become constant at $0.5$
for\ $R_{\lambda }>20$ (see Ref.\cite{SreenAnt}), whereas at\ $R_{\lambda
}>400$ Antonia {\it et al}. \cite{AntChamSaty} show $\left| S\right| $\
increasing from $0.5$ as $\left| S\right| \simeq 0.8R_{\lambda }^{0.11}$. \
If $\varepsilon $\ and $u_{rms}$\ are known from measurements or from flow
similarity, then $R_{\lambda }=u_{rms}^{2}/\left( \varepsilon \nu /15\right)
^{1/2}$ and hence $\left| S\right| $ is known such that (\ref{similarity})
can determine $\left\langle \nu ^{2}\left( \nabla ^{2}{\bf u}\right) {\bf %
\cdot }\left( \nabla ^{2}{\bf u}\right) \right\rangle $. The term $-\left(
1/2\right) \partial _{t}\varepsilon $\ can be neglected if it is not
evaluated. \ The DNS data of Herring and Kerr\cite{HerrKerr}\cite{Kerr85}
suggest that $\left| S\right| \simeq R_{\lambda }/5$ for $R_{\lambda }<1$,
then $\left\langle \nu ^{2}\left( \nabla ^{2}{\bf u}\right) {\bf \cdot }%
\left( \nabla ^{2}{\bf u}\right) \right\rangle \simeq 0.06\varepsilon
^{3/2}\nu ^{-1/2}R_{\lambda }$ for $R_{\lambda }<1$.

\subsection{Menu for the variance of pressure-gradient acceleration}

Poison's equation, local homogeneity and local isotropy, but no other
approximations, result in \cite{HillWilczak} 
\begin{equation}
\left\langle {\bf \nabla }p\cdot {\bf \nabla }p\right\rangle =4\stackrel{%
\infty }{%
\mathrel{\mathop{\int }\limits_{0}}%
}r^{-3}\left[ D_{1111}\left( r\right) +D_{\alpha \alpha \alpha \alpha
}\left( r\right) -6D_{11\beta \beta }\left( r\right) \right] dr,  \label{HW}
\end{equation}
where $D_{1111}\left( r\right) $, $D_{\alpha \alpha \alpha \alpha }\left(
r\right) $, and $D_{11\beta \beta }\left( r\right) $ are components of the
fourth-order velocity structure-function tensor, which is defined by ${\bf D}%
_{ijkl}\left( {\bf r}\right) \equiv \left\langle \left( u_{i}-u_{i}^{\prime
}\right) \left( u_{j}-u_{j}^{\prime }\right) \left( u_{k}-u_{k}^{\prime
}\right) \left( u_{l}-u_{l}^{\prime }\right) \right\rangle $; the $1$-axis
is parallel to the separation vector ${\bf r}$; $\alpha $ and $\beta $
denote the Cartesian axes perpendicular to the $1$-axis. \ Thus, $\alpha $
and $\beta $ are $2$ or $3$; equally valid options under local isotropy are $%
\alpha =\beta $ or $\alpha \neq \beta $.

\qquad There is enough cancellation between the positive and negative parts
of the integrand, i.e., between $r^{-3}\left[ D_{1111}\left( r\right)
+D_{\alpha \alpha \alpha \alpha }\left( r\right) \right] $ and $%
-r^{-3}6D_{11\beta \beta }\left( r\right) $, to make evaluation of the
integral $\stackrel{\infty }{%
\mathrel{\mathop{\int }\limits_{0}}%
}r^{-3}\left[ D_{1111}\left( r\right) +D_{\alpha \alpha \alpha \alpha
}\left( r\right) -6D_{11\beta \beta }\left( r\right) \right] dr$\ difficult
by means of experimental or DNS data\cite{HillWilczak}\cite{HillBoratav}\cite
{NelkinChen98}. \ Hill and Wilczak\cite{HillWilczak} argued that the ratio $%
H_{\chi }\equiv \stackrel{\infty }{%
\mathrel{\mathop{\int }\limits_{0}}%
}r^{-3}\left[ D_{1111}\left( r\right) +D_{\alpha \alpha \alpha \alpha
}\left( r\right) -6D_{11\beta \beta }\left( r\right) \right] dr/\stackrel{%
\infty }{%
\mathrel{\mathop{\int }\limits_{0}}%
}r^{-3}D_{1111}\left( r\right) dr=\left\langle {\bf \nabla }p\cdot {\bf %
\nabla }p\right\rangle /4\stackrel{\infty }{%
\mathrel{\mathop{\int }\limits_{0}}%
}r^{-3}D_{1111}\left( r\right) dr$ is a universal constant at high Reynolds
numbers. \ Universality of $H_{\chi }$ is equivalent to the assertion that $%
\left\langle {\bf \nabla }p\cdot {\bf \nabla }p\right\rangle $ scales with $%
\stackrel{\infty }{%
\mathrel{\mathop{\int }\limits_{0}}%
}r^{-3}D_{1111}\left( r\right) dr$ at high Reynolds numbers. \ Hill and
Wilczak\cite{HillWilczak} pointed out that the utility of determining $%
H_{\chi }$ is that the pressure-gradient variance can then be measured with
a single-wire hot-wire anemometer by means of 
\begin{equation}
\left\langle {\bf \nabla }p\cdot {\bf \nabla }p\right\rangle =4H_{\chi }%
\stackrel{\infty }{%
\mathrel{\mathop{\int }\limits_{0}}%
}r^{-3}D_{1111}\left( r\right) dr.  \label{4Hchiintegral}
\end{equation}
Asymptotic expressions for $\left\langle {\bf \nabla }p\cdot {\bf \nabla }%
p\right\rangle $\ derived from (\ref{4Hchiintegral}) are \cite{Hill01} 
\begin{equation}
\left\langle {\bf \nabla }p\cdot {\bf \nabla }p\right\rangle \simeq
3.1H_{\chi }\varepsilon ^{3/2}\nu ^{-1/2}F^{0.79}\simeq 3.9H_{\chi
}\varepsilon ^{3/2}\nu ^{-1/2}R_{\lambda }^{0.25}\text{ \ for \ }R_{\lambda
}\gtrsim 400,  \label{hiasym}
\end{equation}
where $F\equiv \left\langle \left( \partial _{x_{1}}u_{1}\right)
^{4}\right\rangle /\left\langle \left( \partial _{x_{1}}u_{1}\right)
^{2}\right\rangle ^{2}$, and the right-most expression is obtained from the
data for $F$\ of Antonia {\it et al}.\cite{AntChamSaty}, and, for low
Reynolds numbers\cite{Hill01} 
\begin{equation}
\left\langle {\bf \nabla }p\cdot {\bf \nabla }p\right\rangle \simeq
0.11\varepsilon ^{3/2}\nu ^{-1/2}R_{\lambda }\text{ \ for \ }R_{\lambda
}\lesssim 20.  \label{loasym}
\end{equation}

\qquad Using DNS data, the preferable evaluation of $H_{\chi }$\ is via $%
H_{\chi }=\left\langle {\bf \nabla }p\cdot {\bf \nabla }p\right\rangle /%
\left[ 4\stackrel{\infty }{%
\mathrel{\mathop{\int }\limits_{0}}%
}r^{-3}D_{1111}\left( r\right) dr\right] $ so as to avoid the statistical
uncertainty caused by the cancellations within the integrand of (\ref{HW}).
\ Vedula and Yeung\cite{VedulaYeung} evaluated $H_{\chi }$\ using DNS data
with $R_{\lambda }<230$\ and obtained the variation from $H_{\chi }\simeq
0.55$ at $R_{\lambda }=20$\ to an approach to a constant value of $H_{\chi
}\simeq 0.65$ in the range $80<R_{\lambda }<230$. \ For $H_{\chi }=0.65$, (%
\ref{hiasym}) agrees quantitatively with the DNS data in Table 1 of Gotoh
and Fukayama\cite{GotFuka} for $R_{\lambda }\geq 387$, and (\ref{hiasym}) is
a good approximation for $R_{\lambda }\gtrsim 200$.\cite{Hill01} \ Since the
asymptotic formula (\ref{hiasym}) is thereby verified with $H_{\chi }=0.65$%
,\ the range of validity of $H_{\chi }=0.65$ seems to extend to the highest $%
R_{\lambda }$ for which Antonia {\it et al}.\cite{AntChamSaty} obtained $F$,
namely $R_{\lambda }\simeq 10^{4}$, but it might extend to $R_{\lambda
}=\infty $. \ On that basis, we have the increase from the
low-Reynolds-number asymptote\cite{Hill94} $H_{\chi }=0.36$ as $R_{\lambda
}\longrightarrow 0$, to $H_{\chi }\simeq 0.55$ at $R_{\lambda }=20$, to $%
H_{\chi }\simeq 0.65$ in the range $R_{\lambda }>80$.

\qquad Thus, the menu for calculating $\left\langle {\bf \nabla }p\cdot {\bf %
\nabla }p\right\rangle $\ is (\ref{HW}), or (\ref{4Hchiintegral}) with the
empirically known values of $H_{\chi }$ as a function of $R_{\lambda }$, or (%
\ref{hiasym})-(\ref{loasym}) with the DNS data of Vedula and Yeung\cite
{VedulaYeung} and Gotoh and Fukayama\cite{GotFuka} used in the range $%
20\gtrsim R_{\lambda }\gtrsim 400$.

\section{ISOTROPIC TURBULENCE}

\qquad For isotropic turbulence there is a one-to-one relationship between $%
R_{{\bf a}}$\ and $R_{\lambda }$; it is shown in Fig. 1 with $R_{\nabla p}$
included. \ Fig. 1 was obtained from (\ref{alternative}) by use of (\ref
{hiasym})-(\ref{loasym}) and (\ref{similarity}) and the discussions
following those equations. \ That is: for $R_{\lambda }\lesssim 1$, $%
R_{\nabla p}\simeq \left( 12R_{\lambda }/35\left| S\right| \right)
^{1/2}\simeq \left( 12/7\right) ^{1/2}\simeq 1.3$; for $1\lesssim R_{\lambda
}\lesssim 20$, $R_{\nabla p}\simeq \left( 12R_{\lambda }/35\left| S\right|
\right) ^{1/2}$ with data of Kerr\cite{Kerr85} for $\left| S\right| $; for $%
20\lesssim R_{\lambda }\lesssim 400$, $R_{\nabla p}\simeq 2.6\left(
\left\langle {\bf \nabla }p\cdot {\bf \nabla }p\right\rangle /\varepsilon
^{3/2}\nu ^{-1/2}\right) ^{1/2}$ with data tabulated by Vedula and Yeung\cite
{VedulaYeung} and Gotoh and Fukayama\cite{GotFuka}; for $R_{\lambda }\gtrsim
400$, $R_{\nabla p}\simeq \left[ \left( 3.1H_{\chi }\varepsilon ^{3/2}\nu
^{-1/2}F^{0.79}\right) /\left( 0.3\varepsilon ^{3/2}\nu ^{-1/2}\left|
S\right| \right) \right] ^{1/2}\simeq 3.3R_{\lambda }^{0.07}$; where the
above mentioned $R_{\lambda }$\ variation of $F$\ and $\left| S\right| $\ of
Ref.\cite{AntChamSaty} was used; finally, $R_{{\bf a}}=\left( 1+R_{{\bf %
\nabla }p}^{2}\right) ^{1/2}$. \ In Fig. 1, $R_{\nabla p}\simeq
3.3R_{\lambda }^{0.07}$ is extended to $R_{\lambda }=100$\ to show, by
comparison with the data of Vedula and Yeung and Gotoh and Fukayama, that it
is a good approximation for $R_{\lambda }\gtrsim 200$. \ From $R_{\lambda
}=1 $ to $10^{4}$\ in Fig. 1, $R_{\nabla p}$\ changes by one decade and $R_{%
{\bf a}}$ less so. \ Fig. 1 is based on nearly isotropic data only for $%
R_{\lambda }\lesssim 400$; for $R_{\lambda }\gtrsim 400$ the data used for $%
F $ in (\ref{hiasym}) and $S$ in (\ref{similarity})\ is that of Antonia {\it %
et al}.\cite{AntChamSaty} which consists, in part, of atmospheric surface
layer data at $R_{\lambda }\gtrsim 2000$. \ The assumption is that for $%
R_{\lambda }\gtrsim 400$ the turbulence is locally isotropic and that the
relationships of $F$ and $S$\ to $R_{\lambda }$ measured by Antonia {\it et
al}.\cite{AntChamSaty} do not differ significantly from what they would be
for isotropic turbulence. \ Some support for the assumption is that the
asymptote $R_{\nabla p}\simeq 3.3R_{\lambda }^{0.07}$\ in Fig. 1 agrees with
the nearly isotropic data to $R_{\lambda }\simeq 200$. \ However, futher
confirmation must await DNS and experiments on nearly isotropic turbulence
at higher Reynolds numbers than have been attained to date.

\subsection{Relationship to recent data}

\qquad The DNS data of Vedula and Yeung\cite{VedulaYeung}, Gotoh and Rogallo 
\cite{Gotoh}, and Gotoh and Fukayama\cite{GotFuka} are in the range of $%
R_{\lambda }$\ where $\left| S\right| \simeq 0.5$. \ Since their DNS are
steady state, (\ref{similarity}) gives $\left\langle \nu ^{2}\left( \nabla
^{2}{\bf u}\right) {\bf \cdot }\left( \nabla ^{2}{\bf u}\right)
\right\rangle \simeq 0.15\varepsilon ^{3/2}\nu ^{-1/2}$. \ In their Fig. 1,
Vedula and Yeung\cite{VedulaYeung} show a ratio that they call $\zeta $,
which equals $R_{{\bf \nabla }p}^{2}$, as well as a quantity $a_{0}^{\left(
I\right) }\equiv \left\langle {\bf a}\cdot {\bf a}\right\rangle /\left(
3\varepsilon ^{3/2}\nu ^{-1/2}\right) \simeq 0.05R_{{\bf a}}^{2}$, where $%
\left\langle \nu ^{2}\left( \nabla ^{2}{\bf u}\right) {\bf \cdot }\left(
\nabla ^{2}{\bf u}\right) \right\rangle \simeq 0.15\varepsilon ^{3/2}\nu
^{-1/2}$ was used; they graphed both $\zeta $ and $a_{0}^{\left( I\right) }$
versus $R_{\lambda }$.\ \ Similarly, Gotoh and Rogallo\cite{Gotoh} and Gotoh
and Fukayama\cite{GotFuka} show $F_{{\bf \nabla }p}=3a_{0}^{\left( I\right)
}\simeq 0.15R_{{\bf a}}^{2}$ versus $R_{\lambda }$\ in their Figs. 1 and 2.
\ Therefore, the above-mentioned graphs show $R_{{\bf \nabla }p}^{2}$ and $%
R_{{\bf a}}^{2}$\ as $R_{\lambda }$\ varies. \ Reversing the role of
ordinates and abscissas in their graphs and in Fig. 1, the graphs show $%
R_{\lambda }$\ for nearly isotropic turbulence as the universal Reynolds
number $R_{{\bf a}}$ varies.

\section{ANISOTROPIC TURBULENCE}

\qquad For anisotropic turbulence, the value of $\left\langle
u_{1}^{2}\right\rangle $\ depends, by definition of anisotropy, on the
direction of ${\bf r}$. \ Thus, $R_{\lambda }$\ does also. \ The appearance
of a statistic graphed versus $R_{\lambda }$\ can change depending on which
velocity component is used to calculate $R_{\lambda }$. \ For many flow
geometries, both $R_{{\bf \nabla }p}$ and $R_{{\bf a}}$ increase as $%
R_{\lambda }$ increases. \ It is appropriate to change perspective: \ The
nonuniversal abscissa $R_{\lambda }$ typically, but not necessarily,
increases as the universal abscissas $R_{{\bf \nabla }p}$\ and $R_{{\bf a}}$%
\ increase.

\qquad The turbulent flow in a cylinder driven by counter rotating blades
has several modes of large-scale structure\cite{Zocchietal}\cite{VothPhD}
and is anisotropic at large scales, but local isotropy is approached\cite
{LaPorta} at high Reynolds numbers. \ Belin {\it et al}.\cite{Belinetal}
measured $F$\ in the flow between counter-rotating blades; their $F$\
graphed with $R_{\lambda }$\ on the abscissa\ has a maximum and minimum,
shown here in Fig. 2a. \ That behaviour of $F$\ has been called
controversial \cite{Nelkin00} because it has not been previously observed,
and was not observed by Pearson and Antonia\cite{PearsonAnton} who obtained $%
F$\ in the range of $R_{\lambda }$\ where Belin {\it et al}. observe the
maximum and minimum. \ The Belin {\it et al}.\cite{Belinetal} data for $F$\
is thus a good test of whether or not $R_{{\bf a}}$\ produces a different
organization of a small-scale statistic than does $R_{\lambda }$. \ Only the
single lowest Reynolds-number datum of Belin {\it et al}. is below $%
R_{\lambda }=200$. \ Figure 1 shows that (\ref{hiasym}) is accurate for $%
R_{\lambda }=200$. \ From the menus we therefore select (\ref{similarity})
and (\ref{hiasym}) and use $H_{\chi }=0.65$; then $R_{{\bf \nabla }p}=\left(
2.0F^{0.79}/0.3\left| S\right| \right) ^{1/2}$. \ Thus their data for $F$
and $S$ are used to calculate $R_{{\bf \nabla }p}$, then $R_{{\bf a}}$\ is
obtained from (\ref{definitions}), although $R_{{\bf a}}\simeq R_{{\bf %
\nabla }p}$\ in this case. \ In Fig. 2a the data of Belin {\it et al}. for $F
$\ versus $R_{\lambda }$\ is shown using distinct symbols in several ranges
of $R_{\lambda }$. \ Figure 2b shows their $F$\ versus $R_{{\bf a}}$\ using
the same symbols for the same data points. \ The maximum and minimum in Fig.
2a is absent in Fig.2b, and the symbols are significantly rearranged in Fig.
2b relative to Fig. 2a. Figure 2b is not just $\left( F^{0.79}\right) ^{1/2}$
vs. $F$ because that would be a straight line. \ The data is scattered in
Fig. 2b because of the scatter in $S$. \ Their $S$-values at nearly the same
value of $R_{\lambda }$\ vary by as much as 28\%. \ The important point is
shown by comparing Fig. 2a with 2b: \ The use of $R_{{\bf a}}$\ on the
abscissa rather than $R_{\lambda }$\ can give a significantly different
distribution of the data of a small-scale statistic.

\section{Summary and comment}

\qquad As defined in (\ref{definitions}), $R_{{\bf a}}$\ is a Reynolds
number; it is the ratio of rms fluid-particle acceleration to rms viscous
acceleration; it is composed of statistics of the small scales of
turbulence; it can be used as a universal abscissa for judging the
universality of turbulence statistics. \ Although $R_{{\bf \nabla }p}$ is
not strictly a Reynolds number, it can also be used as a universal abscissa.
\ At high Reynolds numbers, $R_{{\bf \nabla }p}\simeq R_{{\bf a}}$. \ $R_{%
{\bf \nabla }p}$ can be evaluated with single-wire hot-wire anemometry, and
by flow similarity; use of (\ref{alternative}) then determines $R_{{\bf a}}$%
. \ If one chooses to measure $S$\ for the purpose of obtaining the viscous
acceleration variance from (\ref{similarity}), then accurate values of $S$\
require many samples and high spatial resolution. \ This does not require an
enormous data set because the data can be stored as a histogram having one
bin for each of the digitizer's distinct outputs with a bin for digitizer
overflow and one for underflow.

\qquad Models of the small-scale statistics of turbulence should be
expressed in terms of universal attributes instead of in terms of $%
R_{\lambda }$. \ For example, in Table II of Belin {\it et al}.\cite
{Belinetal}, the model by Pullin and Saffman\cite{PullinSaff} is in good
agreement with data when judged in terms of power laws between derivative
moments, but it is in relatively poor agreement with data when judged in
terms of power laws between normalized derivative moments and $R_{\lambda }$%
. \ $R_{\lambda }$ can be specific to the flow geometry.

\begin{acknowledgement}
The author thanks the organizers of the Hydrodynamics Turbulence Program
held at the Institute for Theoretical Physics, UCSB, whereby this research
was supported in part by the National Science Foundation under grant number
PHY94-07194. \ This work was partially supported by ONR Contract No.
N00014-96-F-0011.
\end{acknowledgement}

\bigskip

LIST OF FIGURES

Fig. \ 1 Relation of\ $R_{{\bf a}}$\ (upper symbols) and $R_{\nabla p}$\
(lower symbols) to $R_{\lambda }$ for isotropic turbulence. \ Data of Refs. 
\cite{Kerr85}, \cite{VedulaYeung}, \cite{GotFuka} produce the pluses,
asterisks, and triangles, respectively. \ Solid lines are the low- and
high-Reynolds number asymptotes.

Fig. 2 \ a) The velocity-derivative flatness of Belin {\it et al}.\cite
{Belinetal} versus $R_{\lambda }$\ with different symbols for several ranges
of $R_{\lambda }$; b) the same data and symbols for each point, but versus $%
R_{{\bf a}}$.

\end{document}